\documentstyle[prl,aps,preprint,epsf]{revtex}
\newlength{\figwidth}
\title{Spectral statistics of chaotic systems with a point-like scatterer}
\author{E. Bogomolny, P. Leboeuf and C. Schmit}
\address{Laboratoire de Physique Th\'eorique et Mod\`eles Statistiques
\footnote{Unit\'e de recherche de l'Universit\'e de Paris XI associ\'ee au 
  CNRS}, B\^at. 100, \\ Universit\'e de Paris-Sud, 91405 Orsay Cedex, France}
\begin{document}
\maketitle {\begin{abstract} The statistical properties of a Hamiltonian $H_0$
perturbed by a localized scatterer are considered. We prove that when $H_0$
describes a bounded chaotic motion, the universal part of the spectral
statistics are not changed by the perturbation. This is done first within the 
random matrix model. Then it is shown by semiclassical techniques that 
the result is due to a cancellation between diagonal diffractive and 
off-diagonal periodic-diffractive contributions. The compensation is a very
general phenomenon encoding the semiclassical content of the optical theorem.
\end{abstract}}
{}
\narrowtext

In quantum systems, the chaotic or disordered nature of the classical motion
is reflected in the statistical properties of the high lying eigenvalues and
eigenvectors. For example, the spectral statistics of ballistic cavities are
universal for energy ranges that are small compared to the inverse time of
flight through the system. These universal properties are well described by 
random matrix theory (RMT)  \cite{mehta,boh}.

Consider now a perturbation imposed to a chaotic system. On the classical
side, the dynamics is structurally stable, in the sense that a generic smooth
perturbation leaves the dynamics chaotic. We are interested in the quantum
mechanical effects of a particular class of perturbations that are
non-classical. If the unperturbed motion is described by the Hamiltonian $H_0$
acting in an $N$-dimensional Hilbert space, we consider Hamiltonians of the
form
\begin{equation}\label{1}
H = H_0 + \lambda N | v \rangle \langle v | \ ,
\end{equation}
where $|v \rangle$ is a fixed vector. $N$ is included in the perturbation
for future convenience. The eigenvalues $\{\omega_i\}$ of $H$ satisfy the 
equation
\begin{equation}\label{2}
\sum_k \frac{|v_k|^2}{\omega-\epsilon_k} = \frac{1}{\lambda N} \ ,
\end{equation}
with $\{\epsilon_k\}$ the eigenvalues of $H_0$ and 
$v_k = \langle \varphi_k |v \rangle$ the amplitudes of $| v \rangle$ 
in the eigenbasis of $H_0$.

Rank-one perturbations like in Eq.(\ref{1}-\ref{2}) appear in several
contexts. The most common one is when a local short-range impurity or point
scatterer is added to the system \cite{alb}. The physical consequences of such
a perturbation were studied for Fermi gases \cite{km,and}, in the context of
RMT \cite{am} and for ballistic motion of particles in regular \cite{seb} and
chaotic \cite{sie} cavities. Another context is in the physics of many body
problems. In the basis of $H_0$, the perturbation simply reads $v_k v^*_l$ and
is said to be separable. In a mean field approach, it is the simplest model
leading to collective modes via the residual interaction \cite{bro}. Although
our results are general, we adopt for simplicity the language of the localized
point scatterer.

A local perturbation is purely wave-mechanical. For a system with $f$ degrees
of freedom, it represents a modification of the dynamics in a volume $\propto
(2 \pi \hbar)^f$ in phase space, that tends to zero in the semiclassical
limit. For example, the addition of a point scatterer in a ballistic cavity
leaves invariant the classical motion while quantum mechanically it induces
wave effects like diffraction. The modifications of the eigenvalues by the
perturbation are described by Eq.(\ref{2}). A statistical analysis of the
perturbed spectrum was done for the energy levels of a regular integrable
cavity with a point scatterer in \cite{seb,bgs}. It was demonstrated that a
short range repulsion between the eigenvalues, different from RMT, is induced
by the perturbation, thus considerably modifying the initial Poisson
distribution. More recently, M. Sieber \cite{sie} has studied, using
semiclassical techniques, the modifications by a point scatterer of the
spectral statistics of chaotic systems. He showed that diffractive orbits
produce finite contributions which may induce deviations with respect to the
random matrix model. Whether this deviation really exists for chaotic systems,
or on the contrary if there are other (non-diagonal) semiclassical
contributions that cancel the purely diffractive terms is the question we
answer here.

We prove by two different approaches, namely a purely statistical model and a
semiclassical calculation, that a local perturbation produces no deviations
with respect to RMT. At the first place, assuming that the unperturbed
eigenvalues and eigenvectors components in Eq.(\ref{2}) are distributed
according to RMT, i.e. their joint probability density are given by the
standard formulae \cite{mehta,boh}
\begin{equation} \label{3}
P(\{\epsilon_k\}) \propto \prod_{i>j} |\epsilon_i - \epsilon_j|^\beta \ ,
\end{equation}
and 
\begin{equation} \label{4}
P(\{v_k\}) = \prod_i \left(\frac{\beta N}{2 \pi}\right)^{1-\beta/2} 
\exp (- \beta N |v_i|^2 /2)
\ ,
\end{equation}
we show that the joint probability density for the perturbed
eigenvalues is exactly the same as the distribution of the unperturbed ones
\begin{equation} \label{5}
P(\{\omega_k\}) \propto \prod_{i>j} |\omega_i - \omega_j|^\beta \ .
\end{equation}
Here, $\beta=1$ (resp. $2$) for systems with (resp. without) time-reversal
symmetry. In the second place, and to complete the analysis, a semiclassical
calculation of the spectral form factor is considered. It is expressed as a
double sum over all the periodic and diffractive orbits of the system. The
latter are orbits that are scattered by the perturbation. In \cite{sie} the
diagonal contribution of the diffractive orbits was obtained (cf Eq.({13})
below). We compute the off-diagonal contribution coming from the interference
of periodic and diffractive orbits, and find that this contribution exactly
cancels the diagonal diffractive term. We thus recover the statistics of RMT.
The basic physical ingredient at the basis of this cancellation is the
unitarity of quantum scattering processes, i.e. conservation of the flux
scattered by the impurity. Although our semiclassical result is less general
than Eq.(\ref{5}) -- it is only valid for the short time behavior of a
two-point function --, it applies to systems with arbitrary diffraction
coefficient not expressible as a Hamiltonian of the form (\ref{1}). 

In chaotic and disordered system the local universal fluctuations of the
spectrum are essentially related to the properties of the Jacobian (\ref{3}).
We ignore here problems related to the confinement of the eigenvalues, that
are of minor importance for our purpose. We start the proof of Eq.(\ref{5})
from the joint distribution function of both the old and new eigenvalues,
obtained in Ref. \cite{am}
$$
P(\{\epsilon_i\},\{\omega_j\}) \propto  \frac{\prod_{i>j}
(\epsilon_i - \epsilon_j) (\omega_i - \omega_j)}{\prod_{i,j}
|\epsilon_i - \omega_j|^{1-\beta/2}} {\rm e}^{-\rho \sum_i 
(\omega_i - \epsilon_i)} \ ,
$$
with $\rho = \beta/2 \lambda$. We restrict for simplicity to $\lambda>0$
($\lambda <0$ is treated in the same manner). Eq.(\ref{2}) imposes the
restrictions $\epsilon_i \leq \omega_i \leq \epsilon_{i+1}$ (trapping). The
distribution for the perturbed eigenvalues, $\omega_i$, is then defined as
\begin{eqnarray}\label{7}
P (\{\omega_i\}) &=& \int_{-\infty}^{\omega_1} {\rm d}\epsilon_1 
\int_{\omega_1}^{\omega_2} {\rm d}\epsilon_2 \ldots 
\int_{\omega_{N-1}}^{\omega_N} {\rm d}\epsilon_N P(\{\epsilon_i\},\{\omega_j\})
\nonumber \\
&\propto& {\rm e}^{- \rho \sum_i \omega_i} \prod_{i>j} (\omega_i - \omega_j)
\ W(\beta,\rho) \ , 
\end{eqnarray}
with 
$$
W(\beta,\rho) = \int_{-\infty}^{\omega_1} \frac{{\rm e}^{\rho \epsilon_1} 
{\rm d}\epsilon_1}{F(\epsilon_1)} \ldots \int_{\omega_{N-1}}^{\omega_N} 
\frac{{\rm e}^{\rho \epsilon_N} {\rm d} \epsilon_N}{F(\epsilon_N)}
\prod_{i>j} (\epsilon_i - \epsilon_j ) \ ,
$$
and $F(\epsilon)=\prod_j |\epsilon - \omega_j|^{1-\beta/2}$. 
Using the standard expression for the Vandermonde determinant in $W$ 
and integrating term by term one obtains 
\begin{equation} \label{8}
W(\beta,\rho) = \det \left[ I_j^{(i-1)} \right]_{i,j=1,\ldots, N} \ ,
\end{equation}
where  $I_j^{(i)}= \partial_\rho^i I_j$ is the i-$th$ derivative with
respect to $\rho$ of 
\begin{equation} \label{9}
I_j = I_j^{(0)} = \int_{\omega_{j-1}}^{\omega_j} 
\frac{{\rm e}^{\rho\epsilon} {\rm d}\epsilon}{F(\epsilon)} \ .
\end{equation}
For $j=1$, $\omega_{j-1} = -\infty$. 

It is straightforward to check that the $I_j$'s satisfy, for any $j$, the 
following differential equation \cite{ord}
\begin{equation}\label{10}
\left[ \prod_i (\partial_\rho - \omega_i) + \frac{\beta}{2 \rho}
\sum_i \prod_{j(\neq i)} (\partial_\rho - \omega_j) \right] I_j = 0 \ .
\end{equation}
This differential equation allows to write 
$$
I_j^{(N)} = \sum_{i=0}^{N-1} a_i I_j^{(i)} \ ,
$$
with  some coefficients $a_i$. $W(\beta,\rho)$ as defined in Eq.(\ref{8}) 
is the Wronskian of this equation. It then follows that 
\begin{equation} \label{11}
\partial_\rho W = a_{N-1} \ W \ ,
\end{equation}
where in our case $a_{N-1} = \sum_i \omega_i - \beta N /2\rho$. Integration of
Eq.(\ref{11}) leads to
$$W(\beta,\rho) = \frac{W_0}{\rho^{\beta N/2}}\exp(\rho \sum_i \omega_i).$$ 
When this result is replaced in Eq.(\ref{7}) one gets 
\begin{equation} \label{11p}
P(\{\omega_i \}) \propto W_0 \prod_{i>j} (\omega_i - \omega_j) \ .
\end{equation}
$W_0$ is an integration factor, independent of $\rho$.
We compute it from the asymptotic behavior of $I_j^{(i)}$ when 
$\rho \rightarrow \infty$.
In this  limit the integral in Eq.(\ref{9}) may be evaluated explicitly
$$
\lim_{\rho\rightarrow\infty} I_j \propto \frac{{\rm e}^{\rho \omega_j}}
{\rho^{\beta/2} \prod_{i\neq j} |\omega_i - \omega_j|^{1-\beta/2}} \ .
$$
To leading order, $I_j^{(i)} =\omega_j^i I_j$ and  from Eq.(\ref{8}) one
gets
$$
W_0 \propto \prod_{i>j} \frac{|\omega_i - \omega_j|^{\beta}}{(\omega_i -
  \omega_j)} \ .
$$
From this equation and Eq.(\ref{11p}) we recover the random matrix
distribution function Eq.(\ref{5}). A chaotic system coupled to the
environment through a one-channel antenna has been considered in \cite{sts}.
The model is equivalent to Eq.(\ref{2}) but with imaginary $\lambda$.
For $\lambda \rightarrow \infty$ the imaginary part of the perturbed energies 
is small and Eq.(\ref{5}) is obtained. Our method, which takes explicit care
of the trapping problem, allows to prove this result for arbitrary $\lambda$.

In real physical systems agreement with random matrix theory is observed in a
limited range. This universal behavior concerns correlations over energy
ranges that are small compared to $h/T_{\rm min}$, with $T_{\rm min}$ the
typical period of the shortest periodic orbit. The above random matrix
calculation establishes that the universal part of the spectrum is not changed
by the presence of the scatterer. On the other hand, the non-universal
behavior of the correlation functions occurring at scales of the order of, or
larger than, $h/T_{\rm min}$ are modified by the scattering center, since new
diffractive orbits are introduced \cite{vwr,ps}.

Let us now turn to a semiclassical treatment of the spectral correlations.
These are based on trace formulae expansions of the density of states $d
(\omega) = \sum_k \delta (\omega - \omega_k)$, written as a sum of smoothed 
plus oscillatory terms $d = {\bar d} + d^{\rm (osc)}$. We characterize the
correlations by the spectral form factor defined as
\begin{equation} \label{12}
K(\tau) = \int_{-\infty}^{\infty} \! \frac{{\rm d} \eta}{\bar{d}} \;
\left\langle d^{\text{(osc)}}\left( E + \frac{\eta}{2} \right)
             d^{\text{(osc)}}\left( E - \frac{\eta}{2} \right)
\right\rangle
\; \exp\left( 2 \pi i \eta \tau \bar{d} \right) \; .
\end{equation}
The average indicated by brackets is taken over an energy window containing
many quantum levels but whose size is small compared to $E$. We again consider
a fully chaotic system with a point-like scatterer. In the geometrical theory
of diffraction $d^{\rm (osc)} = d^{\rm (osc)}_p + d^{\rm (osc)}_d$, where
$d^{\rm (osc)}_p$ and $d^{\rm (osc)}_d$ are expressed as interferent sums
over periodic and diffractive orbits, respectively \cite{vwr,ps}. 
\begin{equation}
 d^{\rm (osc)}_{p,d}(E)=\sum_{p,d}A_{p,d}
 \exp (i\frac{S_{p,d}(E)}{\hbar}-i\frac{\pi}{2}\mu_{p,d}),
\end{equation} 
with
\begin{eqnarray}\label{15}
&& A_p = \frac{T_p}{2 \pi \hbar |\det(M_p - 1)|^{1/2}} \nonumber \\
&& A_d = \frac{T_d \, {\cal D}(\vec{n},\vec{n}') \,{\rm e}^{-
i\pi(f+1)/4} |\det N|^{1/2}}{4 \pi \hbar k (2\pi\hbar)^{(f-1)/2}} \ . 
\nonumber
\end{eqnarray}
$S_{p,d}(E)$ is the action of the periodic (resp. diffractive) orbits,
$T_{p,d}$ denotes their period, $M_p$ is the monodromy matrix of the periodic
orbit, $N$ is the matrix 
$N_{ij}=\partial^2 S_d/\partial y_i \, \partial y_j$ (where $\vec{y}$ are
coordinates orthogonal to the diffractive trajectory). ${\cal
D}(\vec{n},\vec{n}')$ is the scattering amplitude of the scattering center
located at $\vec{x_0}$ with incoming $\vec{n}$ and outgoing $\vec{n}'$
directions, defined in terms of the perturbed ($G$) and unperturbed ($G_0$)
Green's functions by the relation
$$
G(\vec{x}, \vec{x}')=G_0(\vec{x}, \vec{x}')+
\frac{\hbar^2}{2m}G_0(\vec{x}, \vec{x_0}){\cal
D}(\vec{n},\vec{n}')G_0(\vec{x_0}, \vec{x}') \ .
$$

Using the properties of the periodic orbits of chaotic systems, the diagonal
contribution of $d^{\rm (osc)}_p$ in Eq.(\ref{12}) gives the short-time random
matrix result $K_p (\tau) = (2/\beta) \tau$ \cite{ber}. The one-scattering
contribution of the diffractive orbits in the same approximation is \cite{sie}
\begin{equation}\label{13}
K_d (\tau) = \frac{\tau^2}{8 \beta \pi^2} \left( \frac{k}{2\pi}\right)
^{2 f - 4} \ \sigma \ ,
\end{equation}
with $k$ the modulus of the wave vector at the impurity and $\sigma$ its total
cross section
\begin{equation}
\sigma = \int |{\cal D}({\vec n},{\vec n}')|^2 {\rm d}\Omega \ {\rm d}\Omega' 
\ .
\end{equation}
(${\rm d}\Omega$ is the solid angle element). For simplicity, we restrict the
calculations to one scattering event (multiple scattering may be considered
likewise).

Our purpose is to compute the off-diagonal cross-term coming from the product
of $d^{\rm (osc)}_p$ and $d^{\rm (osc)}_d$ in Eq.(\ref{12}). The 
semiclassical expression for this contribution is 
\begin{equation}\label{14}
K_{pd}(\tau) = \frac{2 \pi \hbar}{\bar{d}} \langle \sum_{p,d} A_p  A_d^* \;
\exp(i (S_p - S_d)/\hbar)
 \ \delta\left(T-\frac{T_p + T_d}{2}\right) + c.c. \rangle \; .
\end{equation}
After energy smoothing, $K_{pd}$ has significant contributions only from
orbits with close actions $S_p \approx S_d$ (having therefore approximately
the same period). Pairs of orbits satisfying this condition may be constructed
by considering the neighborhood of the forward scattering orbits. To each
periodic orbit passing nearby the scatterer ${\cal O}$ we associate an
``almost periodic'' diffractive orbit that is similar to the periodic orbit
but comes back to ${\cal O}$ with a slightly different momentum. In
Eq.(\ref{14}) the double sum now involves all the possible pairs of
trajectories constructed this way. Consider a surface of section that includes
${\cal O}$ and is transversal to the momentum of the periodic orbit when it
comes nearby ${\cal O}$. Let coordinates measured from ${\cal O}$ and momenta
in the plane be denoted by $(\vec{q},\vec{p})$. Consider all the periodic
orbits of period $T$ that cut the section through a differential element ${\rm
d}^{f-1} q \, {\rm d}^{f-1} p$ located at a distance $\vec{q}$ from ${\cal
O}$. The difference of action between these periodic orbits and the
diffractive orbits associated to them as mentioned above is
\begin{equation}\label{spd}
S_p - S_d = -(1/2) \ Q_{i j} \ q_{i} \ q_{j} \ ,
\end{equation}
with 
$$
Q_{i j}=\partial^2 S/\partial q_{i}\, \partial q_{j}+
\partial^2 S/\partial q_{i}'\, \partial q_{j}+
\partial^2 S/\partial q_{i}'\, \partial q_{j}',
$$
and $\vec{q}$  ($\vec{q}~'$) are initial (resp. final) coordinates on the
surface of section. Moreover, one can show that
\begin{equation}
|\det Q| = |\det(M_p - 1)\det N| \cos^2 \theta \ ,
\label{det}
\end{equation}
where $\theta$ is the angle between the normal to the surface of section and
the momentum of the diffractive orbit.

By generalizing arguments used in the derivation of the Hannay - Ozorio de
Almeida sum rule \cite{hannay} one can prove the following sum rule
\begin{equation}\label{15p}
\sum_p \frac{\delta (T - T_p) \ \chi(\vec{q}_p,\vec{p}_p)}{|\det (M_p - 1
  )|}  = \frac{\int {\rm d}^{f-1} q \, {\rm d}^{f-1} p
\ \chi(\vec{q},\vec{p})}{\Sigma} \ ,
\end{equation}
where $\chi(\vec{q},\vec{p})$ is a test function defined on the surface of
section and $(\vec{q}_p,\vec{p}_p)$ are the coordinates of the points at which
the periodic orbit $p$ crosses the surface of section. $\Sigma = \int {\rm
d}^f {\bf x} \, {\rm d}^f {\bf p} \, \delta (E -H({\bf x},{\bf p}))$ is the
total phase-space volume at energy $E$. From Eq.~(\ref{14}), using
Eqs.(\ref{spd}) and (\ref{15p}), we have
\begin{eqnarray}
K_{pd}= &&\frac{{\bar d} \; \tau^2 {\rm e}^{i \pi (f+1)/4}}{\beta k 
(2 \pi \hbar)^{(f-3)/2}\Sigma}\int\sqrt{|\det(M_p -1)| |\det N|} \nonumber \\
&& \times {\cal D}^*
(\vec{n},\vec{n}) {\rm e}^{-(i/2\hbar) Q_{i j} q_{i}  
q_{j}} {\rm d}^{f-1} q \, {\rm d}^{f-1} p + c.c. \nonumber
\end{eqnarray}
Integrating the quadratic form in the exponent, taking into account
Eq.(\ref{det}), using the semiclassical density of states $\bar{d} = \Sigma/(2
\pi \hbar)^f$ and the fact that the differential element for the momenta may
be written ${\rm d}^{f-1} p = (\hbar k)^{f-1} \cos \theta \ {\rm d}\Omega$,
one obtains the final expression
\begin{equation}\label{16}
K_{pd} (\tau) = \frac{\tau^2}{2\pi\beta}\left(\frac{k}{2\pi}\right)^{f-2} 
\int i \left[ {\cal D}^* ({\vec n},{\vec n}) - {\cal D} ({\vec n},{\vec n})
\right] {\rm d}\Omega \ .
\end{equation}
This is the result for the cross-term contribution. Note that it depends only
on ${\cal D} ({\vec n},{\vec n})$, which translates the fact that
in general interference terms between periodic and diffractive orbits can be
large only for the diffraction in the forward direction.

To make contact with Eq.(\ref{13}) we use a general relation valid for the
elastic scattering on a finite range potential. The conservation of the
flux scattered by the scattering center imposes a relation between the
imaginary part of the scattering amplitude and the scattering cross section.
This is the well known optical theorem \cite{lan}, that in $f$ dimensions 
takes the form
$$
i \left[ {\cal D}^* ({\vec n},{\vec n}) - {\cal D} ({\vec n},{\vec n})
\right] = - \frac{1}{4\pi} \left(\frac{k}{2\pi}\right)^{f-2} 
\int |{\cal D} ({\vec n},{\vec n}')|^2 {\rm d} \Omega'.
$$
Combining this relation with Eq.~(\ref{16}) one gets our final result
\begin{equation}\label{17}
K_{pd} (\tau) = - K_{d} (\tau) \ .
\end{equation}
The interference between periodic and diffractive orbits  exactly cancels
the diagonal contribution of the diffractive orbits, Eq.(\ref{13}). We recover
from semiclassical methods, at least for a two-point function and short times,
the RMT result. 

The two basic elements producing the cancellation are the sum rule (\ref{15p})
and the optical theorem. Only the former is characteristic of chaotic systems,
the latter being very general. The present semiclassical results may be
extended by similar methods to multiple scattering events. In a wider context,
it should be mentioned that this is one of the rare cases in which a
calculation of off-diagonal contributions (whose role is essential in
producing the correct result) is done explicitly for chaotic systems.

We have concentrated on the fluctuation properties of eigenvalues of chaotic
systems, and have demonstrated that they are unchanged by a local
perturbation. This applies to high lying states, were the statistical
hypotheses hold. On the opposite extreme, a local perturbation may lead to
important modifications of the properties of the ground state of the system.
Take for example a negative $\lambda$. According to Eq.(\ref{2}), each
perturbed eigenvalue remains trapped by two unperturbed ones, except the ground
state. The energy of the ground state may diminish arbitrarily with increasing
$|\lambda|$ and, as can easily be shown, the associated wavefunction becomes
more and more localized at the impurity. In our considerations we have ignored
the presence of this ``collective'' mode.

The authors are grateful for many useful discussions with O. Bohigas, M.
Saraceno, M. Sieber and U. Smilansky. After completion of this manuscript we
became aware of related semiclassical results obtained by M. Sieber.


\begin{references}
\bibitem{ord}  To compensate the absolute value in the denominator of the
integrand in Eq.(\ref{9}), for each $j$ the factors in the products in
Eq.(\ref{10}) must be ordered. To simplify the notation we ignore here 
the ordering problems.
\end{references}
\end{document}